\documentclass[prd,nofootinbib,floats,aps,twocolumn,tightenlines,superscriptaddress]{revtex4-1}

\usepackage{graphicx}
\usepackage{bbold}
\usepackage{mathtools}
\usepackage{mathrsfs}
\usepackage{hyperref}
\usepackage{commath}
\usepackage{bm}
\usepackage{amsfonts,amsmath,amssymb,amsthm}
\usepackage[usenames,dvipsnames]{xcolor}
\usepackage{subfigure}
\usepackage{dsfont}
\usepackage{enumitem}

\DeclareMathOperator{\sgn}{sgn}

\newcommand{\ii}{\mathrm{i}}

\newcommand{\hb}{\bar h}
\newcommand{\normord}[1]{:\mathrel{#1}:}
\renewcommand*\d[2][]{%
	\mathrm{d}%
	\ifx\relax#1\relax\else
	\rule{-0.02em}{1.5ex}^{#1}\rule{0.08em}{0ex}\!
	\fi
	#2\,
}

\begin{document}

\title{Gravitational wave emission from the CMB and other thermal fields}
	
\author{Petar Simidzija}
\affiliation{Department of Applied Mathematics, University of Waterloo, Waterloo, Ontario, N2L 3G1, Canada}
\affiliation{Institute for Quantum Computing, University of Waterloo, Waterloo, Ontario, N2L 3G1, Canada}

\author{Achim Kempf}
\affiliation{Department of Applied Mathematics, University of Waterloo, Waterloo, Ontario, N2L 3G1, Canada}
\affiliation{Institute for Quantum Computing, University of Waterloo, Waterloo, Ontario, N2L 3G1, Canada}
\affiliation{Department of Physics and Astronomy, University of Waterloo, Waterloo, Ontario, N2L 3G1, Canada}
\affiliation{Perimeter Institute for Theoretical Physics, Waterloo, Ontario N2L 2Y5, Canada}

\author{Eduardo Mart\'in-Mart\'inez}
\affiliation{Department of Applied Mathematics, University of Waterloo, Waterloo, Ontario, N2L 3G1, Canada}
\affiliation{Institute for Quantum Computing, University of Waterloo, Waterloo, Ontario, N2L 3G1, Canada}
\affiliation{Perimeter Institute for Theoretical Physics, Waterloo, Ontario N2L 2Y5, Canada}

\begin{abstract}

We calculate the gravitational wave power density emitted by quantum thermal sources. As particular cases, we calculate the emission of gravitational waves from the cosmic microwave background and from stellar sources. We study how treating gravity classically affects the prediction of the thermal emission. We find that the predicted emitted gravitational wave radiation does not exhibit an ultraviolet divergence, even if gravity is treated classically, as long as the fields describing the thermal source are quantum mechanical.

\end{abstract}
	
\maketitle

\textit{\textbf{Introduction.}}--- The pursuit of a working theory of quantum gravity has motivated much of the important work in mathematical physics over the past decades. Despite these efforts however, no fully satisfactory theory of quantum gravity has yet been developed. Moreover, the severe difficulties encountered in all attempts to formulate a theory of quantum gravity have led some to question whether gravity even necessitates quantization. Indeed strong theoretical arguments have been made both for and against the need to quantize gravity in the conventional sense, if at all \cite{Eppley1977,Page1981,Mattingly2006,Penrose2014,Belenchia2018}. Ultimately however, it will be down to experiments to settle the debate one way or the other~\cite{Hossenfelder2010,Carlip2008}.

Through this turbulent period of attempting to quantize gravity, we can perhaps take some solace, as well as inspiration, from a similarly turbulent time over a century ago, when both experimental and theoretical physicists were struggling to understand the nature of electromagnetism. Just as today our theory of gravity is highly successful in explaining what it was set out to explain --- namely the dynamics of massive objects such as apples, planets and black holes~\cite{Abbott2016} --- Maxwell's formulation of electromagnetism~\cite{Maxwell1881} in the late 19th century was also highly successful in its original purpose: explaining the outcomes of non-thermal experiments involving charges, currents and fields. However the revolution that followed in our understanding of electromagnetism was not inspired by the successes of classical electromagnetism but rather it came about by considering its failures, namely its catastrophic incompatibility with thermodynamics.

This incompatibility arises when one considers the electromagnetic radiation emitted by a black body. While experiments at the time showed that the radiated energy was always finite, the contemporary understanding of electromagnetism in conjunction with thermodynamics predicted that it should be infinite. This prediction has its origins in the equipartition theorem, which states that in thermal equilibrium at temperature $T$, each degree of freedom of a classical system carries a $k_\textsc{b} T$ amount of energy. Thus a thermal state of a classical field ---which even in a finite volume contains an infinite number of high frequency (ultraviolet) modes--- is predicted to have an infinite energy density. This electromagnetic UV catastrophe was only resolved following the revolutionary work of Max Planck, who argued that the high UV modes contribute a negligible amount to the emitted energy in a way that tamed the divergences of the classical theory. The key assumption that led to this conclusion, which Max Planck was reluctantly forced to make, was that the allowed energies of EM modes were not classical and continuous, but rather discrete, i.e. quantized~\cite{Planck1900}. 

With this historical hindsight in mind, a key question arises: if the UV catastrophe led to the logical conclusion that electromagnetism cannot be classical, can we use an analogous argument to conclude the same for gravity? 

Indeed, there are striking similarities between (linearized) gravity and electromagnetism. General relativity predicts (and recent astrophysical observations confirm~\cite{Abbott2016}) that similarly to how oscillating charges produce electromagnetic waves, so do oscillating masses produce gravitational waves (GWs). More precisely, the linearized Einstein equations in the Lorenz gauge read
\begin{equation}\label{eq:GW}
    \Box\hb_{\mu\nu}=-16\pi G T_{\mu\nu}.
\end{equation}
Here the dynamical quantity $\hb_{\mu\nu}$ is a small perturbation to the flat space metric, and it satisfies a wave equation sourced by the stress-energy density $T_{\mu\nu}$ of the matter fields in the Universe. We can now ask the question: if we consider the matter fields sourcing GWs to be in a thermal state, and if we consider the radiated GWs to be classical, would we find, similar to the case of electromagnetism in the late 19th century, a discrepancy between predictions and observations? If so, we could conclude, in the manner of Planck, that gravity cannot be fully classical. 

Before we proceed to answer this question, it is important to emphasize that despite the physical and mathematical similarities between our setup (a thermal source emitting GWs) to the setup which Planck considered (a thermal source emitting EM radiation) there is also a crucial difference. Namely, in Planck's thought experiment, the thermal source and the EM field were both assumed to classical, but this assumption led to a breakdown of the classical description: the UV catastrophe. Planck resolved this contradiction by removing the second assumption, i.e. he proposed that the EM field is quantized. However, viewing the problem with our present hindsight, it would have been equally logical to attempt to resolve the catastrophe by instead removing the first assumption, i.e. to assume that the matter source (and not the EM field) is quantized. Indeed, this adjustment would also have prevented the UV catastrophe: for instance, an array of quantum harmonic oscillators in a thermal state, coupled to a classical EM field, will emit a non-divergent amount of EM radiation.

Fortunately for us however, we will not have to face an analogous dilemma. Namely, precisely because of the work of Planck (and those that followed) we are now fully aware that matter and light are both quantized. Hence in our setup, with a thermal source emitting GWs, we know for certain that the thermal source, whether it be made from massive matter or EM radiation, must be quantized. The only assumption which we have to make is whether the gravitational field is classical or not. We will assume that the gravitational field is classical, compute the power radiated into GWs, and determine whether or not the result is catastrophic.

As we will show, the result is not catastrophic: the amount of energy radiated into classical GWs is finite, as long as we treat the matter fields that source these GWs quantum mechanically. This is perhaps unsurprising when we recall the analogous result discussed above, namely that the amount of energy radiated into classical EM waves by a thermal source of quantum harmonic oscillators, is finite. However, because we wish to treat the thermal source as a quantum field, and because of the known UV divergences that emerge when one upgrades from a grid of oscillators to a field theory, it is prudent to perform the calculation explicitly, and conclusively verify that classical gravity does not suffer from a thermodynamic UV catastrophe\footnote{Notice that the non-linearity, and therefore the self-interaction, of the gravitational interaction provides a means for the ``heat'' to percolate to UV gravitational-wave modes if gravity were classical. In principle this could lead to a UV catastrophe in the gravity sector if gravity were classical, although this mechanism is highly suppressed due to the weakness of the gravitational self-interaction. Analyzing this problem is beyond our scope here, as it will require a careful analysis due the fact that the very notion of how much energy is carried by gravitational waves is highly non-trivial.}.

Perhaps more importantly, the explicit calculation which we will perform will not only tell us that the power radiated by a thermal quantum field into GWs is finite, but it will also give an order of magnitude estimate of its value. This will allow us to estimate the energy loss due to thermal GW emission of any system that can be approximated as a quantum field, such as the surface of a star, or the cosmic microwave background (CMB). We will find that our quantitative prediction for the power radiated from such sources is small enough that it is consistent with the fact that such a universal mechanism of energy loss has not yet been observed.

\vspace{0.5em}
\textit{\textbf{Setup.}}--- Our goal is to estimate the rate of energy emission $\frac{\dif \epsilon}{\dif t}$ from the CMB into GWs. Then, since we know the energy density of the CMB to be
\begin{equation}
\label{eq:CMB_energy}
    \epsilon = \frac{\pi^2}{15}\frac{(k_\textsc{b}T)^4}{(\hbar c)^3},
\end{equation}
a computation of $\frac{\dif \epsilon}{\dif t}$ will allow us to estimate what percentage of its energy the CMB is losing to GWs in a unit time. Note that this simple method of estimating the net rate of energy flow from the CMB to the GWs is only appropriate under the assumption that there is no back-flow of energy from the GWs to the CMB. This assumption will be justified a posteriori. In what follows, we will use natural units $k_\textsc{b}=\hbar=c=1$.

To compute $\frac{\dif \epsilon}{\dif t}$, we need to be able to quantify the energy carried by GWs. However, quantifying the `spacetime-curving' energy carried by spacetime itself is non-trivial, and in general it is not even a meaningful notion. Nevertheless, in the linear regime, it is possible to assign an effective stress-energy density $t_{\mu\nu}$ to a propagating GW perturbation $\hb_{\mu\nu}$. Such an expression can be obtained either geometrically or field theoretically --- in either case the result is~\cite{Maggiore2008}
\begin{equation}
\label{eq:t_mu_nu}
    t_{\mu\nu}=\frac{1}{32\pi G}
    \partial_\mu h_{\alpha\beta}
    \partial_\nu h^{\alpha\beta},
\end{equation}
where $h_{\mu\nu}=\hb_{\mu\nu}-\frac{1}{2} \eta_{\mu\nu}\hb_{\alpha}^{\phantom{\alpha}\alpha}$, and all indices are raised with the flat metric $\eta_{\mu\nu}$. 


There are two important points to note regarding Eq.~\eqref{eq:t_mu_nu}. First, this simple expression is only valid outside of sources. This will not be a problem for our purposes: we will compute the GW power radiated out to a distance $r$ by a finite volume $V$ of characteristic length $L\ll r$ containing the CMB. Then, dividing by $V$ will give us the desired power density $\frac{\dif \epsilon}{\dif t}$. In particular we will find that $\frac{\dif \epsilon}{\dif t}$ is independent of $V$, thus justifying our approach of restricting to a finite volume. 

Second, and more fundamentally, it is only possible to define a meaningful effective stress-energy of a GW on scales larger than its characteristic wavelength, and therefore $t_{\mu\nu}$ only makes sense in a spatially averaged sense~\cite{Maggiore2008}. Since we expect the characteristic wavelength of GWs emitted from a thermal source of temperature $T$ to be of order $\beta=1/T$, our results will only be meaningful if we take $L\gg \beta$. 

Since we are interested in the total power radiated by the volume $V$, we need to integrate the energy density in Eq.~\eqref{eq:GW} over a spherical shell of radius $r\gg L$. If we assume that $V$ is also a sphere, we have
\begin{equation}
\label{eq:dE/dt}
    \frac{\dif \epsilon}{\dif t}
    =
    \frac{4\pi r^2 t_{00} }{V}
    =
    \frac{4\pi r^2}{32\pi GV}  
    \dot h_{\mu\nu}
    \dot h^{\mu\nu},
\end{equation}
where a dot denotes a time derivative. By spherical symmetry, this expression can be evaluated anywhere on the shell $r\gg L$. 

Furthermore, since we are only considering $h_{\mu\nu}$ in the vacuum, it is possible to write it in the \textit{transverse-traceless} (TT) gauge. Indeed, given a plane wave $h_{\mu\nu}$ in the Lorenz gauge, propagating in the $\hat n$ direction, we can write it in the TT gauge as $h_{ij}^\textsc{tt}=\Lambda_{ij}^{\phantom{ij}kl}(\hat n)h_{kl}$, where $\Lambda_{ijkl}(\hat n)$ is a spatial projection tensor (see~\cite{Maggiore2008} for details). In the TT gauge $h_{\mu\nu}^\textsc{tt}=\hb_{\mu\nu}^\textsc{tt}$, and hence Eq.~\eqref{eq:dE/dt} becomes
\begin{equation}
\label{eq:dE/dt_TT}
    \frac{\dif \epsilon}{\dif t}
    =
    \frac{4\pi r^2}{32\pi GV} 
    \Lambda_{ij}^{\phantom{ij}kl}
    \Lambda^{ij}_{\phantom{ij}mn}
    \dot{\bar{h}}_{kl}
    \dot{\bar{h}}^{mn}.
\end{equation}
Here, we have used the fact that far from the source the $\hb_{\mu\nu}$ are essentially plane waves. This form for the radiated power is useful since we can invert Eq.~\eqref{eq:GW} to obtain $\hb_{\mu\nu}(\mathsf{x})=-16\pi G \int\d[4]\mathsf{x}' \mathcal{G}(\mathsf{x}-\mathsf{x}') T_{\mu\nu}(\mathsf{x}')$, where $\mathcal{G}$ is a retarded Green's function. Hence, at large distances from the source we find
\begin{equation}
\label{eq:dE/dt_function_of_T}
    \frac{\dif \epsilon}{\dif t}
    =
    \frac{2G}{V}
    \Lambda_{ij}^{\phantom{ij}kl}
    \Lambda^{ijmn}\!
    \int_V\!\!\d[3]{\bm x'}\!\!
    \int_V\!\!\d[3]{\bm x''}
    \dot T_{kl}(\mathsf{x}')
    \dot T_{mn}(\mathsf{x}''),
\end{equation}
where the stress-energy tensor $T_{ik}(\mathsf{x}')$ is evaluated at the retarded time associated with the source spacetime point $\mathsf{x}'$ and field spacetime point $\mathsf{x}$.

\vspace{0.5em}
\textit{\textbf{Semi-classical gravity.}}--- As it stands, Eq.~\eqref{eq:dE/dt_function_of_T} for the power density emitted by the CMB into GWs is a purely classical expression. However in order to avoid the electromagnetic UV catastrophe (i.e. to ensure that the CMB energy density is finite), we need to ensure that the electromagnetic (EM) field is modeled quantum mechanically. The simplest way one might think to introduce quantum behaviour of the matter fields sourcing gravity would perhaps be to use the formalism of semi-classical gravity (SCG), in which we replace the classical $T_{\mu\nu}$ with its quantum expectation value in a thermal state of temperature $T$. Hence, in the SCG formalism, Eq.~\eqref{eq:dE/dt_function_of_T} takes the form
\begin{equation}
\label{eq:dE/dt_SCG}
    \frac{\dif \epsilon}{\dif t}
    \Big|_\textsc{scg}
    \sim
    \frac{G}{V}
    \int_V\!\!\d[3]{\bm x'}\!\!
    \int_V\!\!\d[3]{\bm x''}
    \langle
    \normord{\dot{\hat{T}}_{kl}(\mathsf{x}')}
    \rangle
    \langle
    \normord{\dot{\hat{T}}_{mn}(\mathsf{x}'')}
    \rangle,
\end{equation}
where for brevity we omit writing constant prefactors and the Lambda tensors. Note that the operators $\hat T_{\mu\nu}$ are normal-ordered in order to ensure that $\langle\normord{\hat T_{\mu\nu}}\rangle$ vanishes in the vacuum state~\cite{Wald1994}.

The expectation values in Eq.~\eqref{eq:dE/dt_SCG} are taken in a thermal state, which by definition is a stationary (time independent) state. Therefore $\partial_t\langle\normord{\hat T_{\mu\nu}}\rangle$ identically vanishes, and so SCG predicts $\frac{\dif \epsilon}{\dif t}=0$. This should not come as a surprise: SCG only accounts for the mean value of $\hat T_{\mu\nu}$ and not its fluctuations, whereas we expect that it is precisely these fluctuations that source GWs. Hence, in order to predict more reliably how much power the CMB emits into GWs, we need a theory that takes into account the quantum fluctuations of the CMB stress-energy. For this we use the formalism of \textit{passive quantum gravity} (PQG).\footnote{In both SCG and PQG the gravitational field remains fully classical, while the matter fields posses varying degrees of quantumness. In particular, the gravitational fluctuations in PQG are \textit{passively} induced by the quantum fluctuations of the sources, rather than \textit{actively} induced by its own quantum fluctuations.}

\vspace{0.5em}
\textit{\textbf{Passive quantum gravity.}}--- PQG has been used extensively to study the effects of stress-energy fluctuations of the gravitational field~\cite{Ford1995,Ford2003,Ford2004,Ford2007}. PQG is closely related to the theory of stochastic gravity~\cite{Hu2008}. While in SCG gravity is sourced by the expectation value of $\hat T_{\mu\nu}$, in PQG we only take the expectation value after we have formally solved for the quantity of interest. Thus if this quantity is nonlinear in $\hat T_{\mu\nu}$, then PQG will give a different result than SCG. For example, in the PQG approach $\frac{\dif \epsilon}{\dif t}$ becomes
\begin{equation}
\label{eq:dE/dt_PQG}
    \frac{\dif \epsilon}{\dif t}
    \Big|_\textsc{pqg}
    \sim
    \frac{G}{V}
    \int_V\!\!\d[3]{\bm x'}\!\!
    \int_V\!\!\d[3]{\bm x''}
    \langle
    \normord{\dot{\hat{T}}_{kl}(\mathsf{x}')}
    \normord{\dot{\hat{T}}_{mn}(\mathsf{x}'')}
    \rangle,
\end{equation}
which manifestly depends on the quantum fluctuations of the stress-energy. Note that in the limit of vanishing stress-energy fluctuations we have $\langle \hat T^2\rangle \rightarrow \langle \hat T\rangle^2$, and Eq.~\eqref{eq:dE/dt_PQG} reduces to the result of SCG given in Eq.~\eqref{eq:dE/dt_SCG}. 

To obtain an order of magnitude estimate of the power radiated by the CMB into GWs, we can safely model the CMB using a massless scalar field $\hat\phi(x)$. The stress-energy is then $\hat T_{\mu\nu}=\partial_\mu\hat\phi\partial_\nu\hat\phi-\frac{1}{2}\eta_{\mu\nu}(\partial_\alpha\hat\phi)^2$. As shown explicitly in Appendix~\ref{app:Wick}, substituting this into Eq.~\eqref{eq:dE/dt_PQG} and using Wick's theorem results in
\begin{align}
\label{eq:dE/dt_PQG_2}
    \frac{\dif \epsilon}{\dif t}
    \Big|_\textsc{pqg}\!
    \sim
    \frac{G}{V}\!\!
    \int_V\!\d[3]{\bm x'}\!\!\!
    \int_V\!\d[3]{\bm x''}\!
    \Big[
    &
    \langle
    \dot{\hat\phi}_k(\mathsf{x}')
    \dot{\hat\phi}_m(\mathsf{x}'')
    \rangle
    \langle
    {\hat\phi}_l(\mathsf{x}')
    {\hat\phi}_n(\mathsf{x}'')
    \rangle
    \notag\\
    &+\text{permutations}
    \Big],
\end{align}
where $\hat\phi_k=\partial\hat\phi/\partial x_k$, and all the omitted terms are of the same form as the first term, but with the time derivatives permuted amongst the four field operators. In particular all of the terms in the above sum contain only two-point correlators between $\mathsf{x}'$ and $\mathsf{x}''$, and not, say, between $\mathsf{x}'$ and itself. Notice that there is no normal ordering in these correlators, and so we anticipate having to deal with zero-point divergences.

To proceed with the calculation of Eq.~\eqref{eq:dE/dt_PQG_2}, we need to first specify the boundary conditions of the field $\hat\phi(\mathsf{x})$ at the boundary of $V$. Note however that in a thermal state of inverse temperature $\beta$, the correlator $\langle \hat\phi(\mathsf{x}')\hat\phi(\mathsf{x}'')\rangle$ will strongly depend on the choice of boundary conditions only if $\bm x'$ or $\bm x''$ are near (within a distance $\beta$ of) the boundary. Further away from the boundary, the correlator is well approximated by its value in free space. Thus, since we are already considering the limit $L\gg\beta$, we can effectively neglect all edge effects in the integrals in Eq.~\eqref{eq:dE/dt_PQG_2} and approximate $\langle \hat\phi(\mathsf{x}')\hat\phi(\mathsf{x}'')\rangle$ by the free space correlator~\cite{Weldon2000}
\begin{equation}
\label{eq:correlator}
    \langle \hat\phi(\mathsf{x}')\hat\phi(\mathsf{x}'')\rangle
    =
    \sum_{n=-\infty}^\infty
    \frac{1}{4\pi^2
    \left[
    \Delta \bm x^2
    -
    (\Delta t+\ii n \beta)^2
    \right]},
\end{equation}
where $\Delta \bm x=|\bm x'-\bm x''|$ and $\Delta t=t'-t''$. Notice that the $\beta$-independent $n=0$ term (which we will call the vacuum (VAC) term) is precisely the vacuum correlator, and is singular for $\Delta\bm x=\pm \Delta t$. Meanwhile, the $\beta$-dependent $n\neq 0$ terms are non-singular, and can be summed up to a finite value for all $\Delta\bm x$ and $\Delta t$. We call these the NO terms, since they are what remains if we normal ordered the fields inside the correlator. Schematically $\langle \hat\phi(\mathsf{x}')\hat\phi(\mathsf{x}'')\rangle = (\text{VAC term})+ (\text{NO terms})$.

As anticipated, upon inserting the field correlator into Eq.~\eqref{eq:dE/dt_PQG_2}, we find that some of the the resulting integrals diverge. The integrands responsible for the divergences are precisely those that contain a VAC term, i.e. they are either of the form $(\text{VAC})\times(\text{VAC})$ or $(\text{VAC})\times(\text{NO})$. The first step towards obtaining a finite result using PQG is to remove the purely VAC terms~\cite{Ford2001,Ford2002,Ford2003}. In other words we are only interested in the relative difference between how much a thermal state radiates and how much the vacuum radiates, which we expect to be zero. (This is analogous to subtracting away the vacuum energy in order to compute the correct energy density of the CMB in Eq.~\eqref{eq:CMB_energy}.) However, even following this vacuum subtraction, the expression for $\frac{\dif \epsilon}{\dif t}$ still contains divergent $(\text{VAC})\times(\text{NO})$ terms, which, being state dependent, cannot be simply subtracted away.

Fortunately, the theory of PQG has a well-defined built-in procedure for dealing with the divergences associated with the the $(\text{VAC})\times(\text{NO})$ terms, which occur generically in all PQG calculations~\cite{Ford2001,Ford2002,Ford2003}. In Appendix~\ref{app:regularization} we show in explicit detail how to regulate these divergences, while here we focus on the main ideas.

To that end, recall that $(\text{VAC})$ represents a higher order derivative of the two-point vacuum correlator, and hence it has a $1/x^n$ type divergence at $\Delta\bm x=\pm \Delta t$. Meanwhile $(\text{NO})$ is a smooth function whose derivatives decay strongly for $|\bm x'-\bm x''|\gg \beta$. Thus our problem is essentially to regularize the integral
\begin{equation}
    \label{eq:simple_int}
    I(L) = 
    \int_{-\frac{L}{2}}^{\frac{L}{2}}
    \frac{f(x)}{x^n}\dif x,
\end{equation}
where $n>1$ is an integer and $f(x)$ and its derivatives decay strongly for $x\gg\beta$.

The above integral $I(L)$ can be defined through an integration by parts procedure introduced in references \cite{Davies1989,Davies1990}: Inserting the identity $\frac{1}{x^n}=\frac{(-1)^{n-1}}{(n-1)!}\frac{\dif^{\,n}}{\dif x^n}\log|x|$ into $I(L)$ and integrating by parts $n$ times, one obtains
\begin{equation}
    \label{eq:simple_int_2}
    I(L) = \frac{-1}{(n-1)!}
    \!\int_{\frac{-L}{2}}^{\frac{L}{2}}
    f^{(n)}(x)\log|x|\dif x
    +
    \left(
    \genfrac{}{}{0pt}{}{\text{boundary}}{\text{terms}}
    \right).
\end{equation}
Since $f(x)$ and its derivatives decay strongly for $x\gg\beta$, we can neglect the boundary terms if $L\gg\beta$. Furthermore, the logarithmic singularity in the remaining integral is integrable, and thus defines a finite value for $I(L)$. Note that the limit $L\gg \beta$, which we originally imposed such that the energy density of GWs be well defined, is also necessary for this regularization procedure to work.

Hence, following regularization, the expression \eqref{eq:dE/dt_PQG_2} for the power density emitted by the CMB into GWs is given by $\frac{\dif\epsilon}{\dif t}\sim\frac{G}{V}\int_V\d[3]{\bm x'}\int_V\d[3]{\bm x''}g(\bm x'-\bm x'')$, where $g(\bm x)$ is an integrable function which decays strongly for $|\bm x|\gg \beta$. It is convenient to approximate this integral as
\begin{equation}
    \label{eq:dE/dt_PQG_3}
    \frac{\dif \epsilon}{\dif t}
    \Big|_\textsc{pqg}\!\!\!
    \sim
    \frac{G}{V}\!\!
    \int_\mathbb{R}\!\!\d[3]{\bm x'}
    e^{\frac{-2|\bm x'|^2}{L^2}}\!\!\!
    \int_\mathbb{R}\!\!\d[3]{\bm x''}
    e^{\frac{-2|\bm x''|^2}{L^2}}
    g(\bm x'-\bm x''),
\end{equation}
where the Gaussian ``smearing" functions pick out the effective domain of integration $V$. We explicitly show the numerical computation of an expression of this form in Appendix~\ref{app:evaluation}. We can however straightforwardly approximate this integral in the limit $L\gg\beta$. Making the substitutions $\bm u = \bm{x'}-\bm{x''}$ and $\bm v = \bm{x'}+\bm{x''}$, we find that
\begin{equation}
    \label{eq:dE/dt_PQG_4}
    \frac{\dif \epsilon}{\dif t}
    \Big|_\textsc{pqg}
    \sim
    \frac{8G}{V}
    \int_\mathbb{R}\d[3]{\bm v}
    e^{\frac{-|\bm v|^2}{L^2}}
    \int_\mathbb{R}\d[3]{\bm u}
    e^{\frac{-|\bm u|^2}{L^2}}
    g(\bm u).
\end{equation}
Note that the first integral is proportional to $L^3\sim V$. Meanwhile in the $L\gg \beta$ limit, the second integral is independent of $L$, since the function $g(\bm u)$ rapidly decays to zero for $|\bm u|> \beta$. Hence, modulo prefactors of order unity, our final expression for $\frac{\dif \epsilon}{\dif t}$ reads
\begin{equation}
    \label{eq:dE/dt_PQG_final}
    \frac{\dif \epsilon}{\dif t}
    \Big|_\textsc{pqg}
    \sim
    \left(\frac{G}{V}\right)V
    =
    \frac{G(k_\textsc{b}T)^7}{\hbar^5 c^8},
\end{equation}
where in the final step we have reinstated the scales $\hbar$, $k_\textsc{b}$, $c$ and $T$ in the only dimensionally consistent manner possible. This expression gives the power density radiated by an electromagnetic field of temperature $T$ into GWs. Interestingly, we contrast the $T^7$ dependence of the this result with the famous $T^4$ dependence of the Stefan-Boltzmann law: thermal radiation into GWs is more sensitive to temperature than thermal radiation into electromagnetic waves.

\vspace{0.5em}
\textit{\textbf{Discussion.}}--- Suppose that the CMB initially has an energy density $\epsilon$ given by Eq.~\eqref{eq:CMB_energy}, and it radiates a small fraction $\delta\ll 1$ of this energy in a time interval $t_\delta$. Then $\frac{\dif \epsilon}{\dif t}\approx\frac{\delta\epsilon}{t_\delta}$ gives
\begin{equation}
\label{eq:rad_time}
    t_\delta\sim
    \frac{\hbar^2c^5}{G(k_\textsc{b}T)^3}
    \delta.
\end{equation}
Notice that if we restrict ourselves to the $\delta\ll 1$ regime in which this equation is valid, then we are also justifying our assumption that the backreaction of the GWs on the CMB can be neglected. 

Notice that, as we would expect, a stronger coupling $G$ between matter and gravity results in the CMB being able to faster radiate its energy into GWs. On the other hand, a ``more quantum'' CMB (higher $\hbar$) takes longer to radiate its energy into GWs. This also agrees with intuition: a ``more quantum'' photon gas in a thermal state has a larger portion of its energy concentrated in low frequency modes, and these slower oscillating modes take longer to radiate away their energy through GWs. Finally, it also makes intuitive sense that a higher temperature CMB would be able to faster radiate away a small fraction of its energy. Indeed (somewhat coincidentally), the time it takes for a Maxwell-Boltzmann gas to radiate away a small fraction of its energy (into electromagnetic radiation) is also proportional to $1/T^3$.

Given that our result passes these intuition checks, let us now use it to answer our principal question: is classical gravity consistent with our thermodynamic observations, or does it, like classical electromagnetism, suffer from a UV catastrophe? 

To answer this, first note that the most precise measurements of the CMB temperature give an average value of $T_0=2.725\,48\pm 0.000\,57$ K~\cite{Fixsen2009}. In order for this value to change by more than its uncertainty, the CMB would have to radiate away at least $0.1\%$ of its energy into GWs. Using Eq.~\eqref{eq:rad_time} with $T=T_0$, we find that this would take roughly $10^{50}$ seconds, a time much larger than the present age of the Universe, $10^{17}$ s. Even if the CMB was radiating GWs at the rate corresponding to $T=3\,000$ K, the temperature of the CMB when radiation first decoupled from matter, it would still take $10^{20}$ years for the CMB to radiate away $0.1\%$ of its present day energy. 


Notice that we have studied the case of gravitational wave emission on a flat background. Our calculation, therefore, describes the thermal emission from the CMB in the form of gravitational waves for timescales much smaller than the Hubble time. Given the large magnitude of time that we find it takes for the CMB to radiate a significant fraction of its energy as GWs, we expect that taking into account a moderate amount of expansion of the universe does not change our conclusions qualitatively. However, it could still be very interesting to go beyond our approximations here, with an analysis of the thermal emission of GWs on a cosmological (i.e. expanding) background spacetime, especially in the very early universe.

We could also use our results to estimate the energy lost by the hottest stars in our galaxy due to GW emission. Admittedly this is only a rough approximation, since, unlike the CMB, the gas inside a star is not necessarily well-modeled by a massless scalar field. In any case there is also no observable energy losses: using equation \eqref{eq:rad_time}, we find that it would take a $40\,000$ K star roughly $10^{29}$ years to lose enough energy through GW emission for us to detect a $0.1\%$ frequency shift in the peak of its electromagnetic spectrum. Therefore, the amount of energy emitted by thermal objects through GWs, which we computed by keeping the gravitational degrees of freedom in our theory fully classical, is not only non-catastrophic, but rather is so small that it is likely to never be observed. 

Finally, while our results do not answer the question of whether gravity requires quantization, it does offer an important insight into this problem. Namely, we have found that while the electromagnetic field was originally quantized in order to avoid the UV catastrophe, this is not necessary for the (linearized) gravitational field. In other words, as long as the matter fields sourcing gravitational waves are treated quantum mechanically, the thermodynamic features of classical gravity remain fully consistent with observations, at least as long as one neglects the (very weak) gravitational self-interaction.

\hspace{0.5em}
\textit{\textbf{Acknowledgements.}}--- P.S. acknowledges the support of the NSERC CGS-M and OGS Scholarships. AK and E.M-M acknowledge support through the NSERC Discovery Program.  E.M.-M. also acknowledges the funding of his Ontario Early Research Award. AK acknowledges support through a Google Faculty Research Award.

\onecolumngrid
\appendix

\section{A variation of Wick's theorem}
\label{app:Wick}

We will now show explicitly how to go from Eq.~\eqref{eq:dE/dt_PQG} to Eq.~\eqref{eq:dE/dt_PQG_2}. We make use of the following variation of Wick's theorem:

\textit{Lemma}: Let $\hat\phi_i(\mathsf{x})$ be a field operator, or a derivative of a field operator. Then
\begin{align}
    \langle
    \normord{\hat\phi_1(\mathsf{x})\hat\phi_2(\mathsf{x})}
    \normord{\hat\phi_3(\mathsf{x}')\hat\phi_4(\mathsf{x}')}
    \rangle
    =
    \langle\normord{\hat\phi_1(\mathsf{x})\hat\phi_2(\mathsf{x})}\rangle
    \langle\normord{\hat\phi_3(\mathsf{x}')\hat\phi_4(\mathsf{x}')}\rangle
    &+
    \langle\hat\phi_1(\mathsf{x})\hat\phi_3(\mathsf{x}')\rangle
    \langle\hat\phi_2(\mathsf{x})\hat\phi_4(\mathsf{x}')\rangle
    \notag\\
    &+
    \langle\hat\phi_1(\mathsf{x})\hat\phi_4(\mathsf{x}')\rangle
    \langle\hat\phi_2(\mathsf{x})\hat\phi_3(\mathsf{x}')\rangle,
\end{align}
where the expectation values are taken in a thermal state.

\textit{Proof}: Using the definition of normal ordering, $\normord{\hat\phi_1\hat\phi_2}=\hat\phi_1\hat\phi_2-\langle\hat\phi_1\hat\phi_2\rangle_0$, where $\langle\cdot\rangle_0$ is the expectation value with respect to the vacuum state, gives
\begin{align}
    \langle
    \normord{\hat\phi_1(\mathsf{x})\hat\phi_2(\mathsf{x})}
    \normord{\hat\phi_3(\mathsf{x}')\hat\phi_4(\mathsf{x}')}
    \rangle
    &=
    \left\langle
    \left(
    \hat\phi_1(\mathsf{x})\hat\phi_2(\mathsf{x})-
    \langle\hat\phi_1(\mathsf{x})\hat\phi_2(\mathsf{x})\rangle_0
    \right)
    \left(
    \hat\phi_3(\mathsf{x}')\hat\phi_4(\mathsf{x}')-
    \langle\hat\phi_3(\mathsf{x}')\hat\phi_4(\mathsf{x}')\rangle_0
    \right)
    \right\rangle
    \notag\\
    &=
    \langle
    \hat\phi_1(\mathsf{x})\hat\phi_2(\mathsf{x})\hat\phi_3(\mathsf{x}')\hat\phi_4(\mathsf{x}')
    \rangle
    -
    \langle\phi_1(\mathsf{x})\hat\phi_2(\mathsf{x})\rangle
    \langle\hat\phi_3(\mathsf{x}')\hat\phi_4(\mathsf{x}')\rangle_0
    \notag\\
    &\hspace{3em}-
    \langle\phi_1(\mathsf{x})\hat\phi_2(\mathsf{x})\rangle_0
    \langle\hat\phi_3(\mathsf{x}')\hat\phi_4(\mathsf{x}')\rangle
    +
    \langle\phi_1(\mathsf{x})\hat\phi_2(\mathsf{x})\rangle_0
    \langle\hat\phi_3(\mathsf{x}')\hat\phi_4(\mathsf{x}')\rangle_0.
\end{align}
Suppose now, for simplicity that $x_0\le x'_0$ (the case $x_0> x'_0$ is treated analogously). Then, applying Wick's theorem for fields~\cite{Peskin1996} on the first term, and cancelling like terms, results in
\begin{align}
    \langle
    \normord{\hat\phi_1(\mathsf{x})\hat\phi_2(\mathsf{x})}
    \normord{\hat\phi_3(\mathsf{x}')\hat\phi_4(\mathsf{x}')}
    \rangle
    =
    &\langle\hat\phi_1(\mathsf{x})\hat\phi_3(\mathsf{x}')\rangle_0
    \langle\hat\phi_2(\mathsf{x})\hat\phi_4(\mathsf{x}')\rangle_0
    +
    \langle\hat\phi_1(\mathsf{x})\hat\phi_4(\mathsf{x}')\rangle_0
    \langle\hat\phi_2(\mathsf{x})\hat\phi_3(\mathsf{x}')\rangle_0
    \notag\\
    &+
    \langle\normord{\hat\phi_1(\mathsf{x})\hat\phi_3(\mathsf{x}')}\rangle
    \langle\hat\phi_2(\mathsf{x})\hat\phi_4(\mathsf{x}')\rangle_0
    +
    \langle\hat\phi_1(\mathsf{x})\hat\phi_3(\mathsf{x}')\rangle_0
    \langle\normord{\hat\phi_2(\mathsf{x})\hat\phi_4(\mathsf{x}')}\rangle
    \notag\\
    &+
    \langle\normord{\hat\phi_1(\mathsf{x})\hat\phi_4(\mathsf{x}')}\rangle
    \langle\hat\phi_2(\mathsf{x})\hat\phi_3(\mathsf{x}')\rangle_0
    +
    \langle\hat\phi_1(\mathsf{x})\hat\phi_4(\mathsf{x}')\rangle_0
    \langle\normord{\hat\phi_2(\mathsf{x})\hat\phi_3(\mathsf{x}')}\rangle
    \notag\\
    &+
    \langle
    \normord{\hat\phi_1(\mathsf{x})\hat\phi_2(\mathsf{x})\hat\phi_3(\mathsf{x}')\hat\phi_4(\mathsf{x}')}
    \rangle.
\end{align}
The result up to now is true for the expectation value in any state. Now, in Ref.~\cite{Ford2004}, starting at Eq. (52), the authors prove that for a \textit{thermal} state the following identity holds:
\begin{align}
    \langle
    \normord{\hat\phi_1(\mathsf{x})\hat\phi_2(\mathsf{x})\hat\phi_3(\mathsf{x}')\hat\phi_4(\mathsf{x}')}
    \rangle
    =
    \langle\normord{\hat\phi_1(\mathsf{x})\hat\phi_2(\mathsf{x})}\rangle
    \langle\normord{\hat\phi_3(\mathsf{x}')\hat\phi_4(\mathsf{x}')}\rangle
    &+
    \langle\normord{\hat\phi_1(\mathsf{x})\hat\phi_3(\mathsf{x}')}\rangle
    \langle\normord{\hat\phi_2(\mathsf{x})\hat\phi_4(\mathsf{x}')}\rangle
    \notag\\
    &+
    \langle\normord{\hat\phi_1(\mathsf{x})\hat\phi_4(\mathsf{x}')}\rangle
    \langle\normord{\hat\phi_2(\mathsf{x})\hat\phi_3(\mathsf{x}')}\rangle.
\end{align}
Combining the previous two expressions, and performing some convenient factoring, one obtains
\begin{align}
    \langle
    \normord{\hat\phi_1(\mathsf{x})\hat\phi_2(\mathsf{x})}
    \normord{\hat\phi_3(\mathsf{x}')\hat\phi_4(\mathsf{x}')}
    \rangle
    =
    &\left(
    \langle\normord{\hat\phi_1(\mathsf{x})\hat\phi_3(\mathsf{x}')}\rangle
    +\langle\hat\phi_1(\mathsf{x})\hat\phi_3(\mathsf{x}')\rangle_0
    \right)
    \left(
    \langle\normord{\hat\phi_2(\mathsf{x})\hat\phi_4(\mathsf{x}')}\rangle
    +\langle\hat\phi_2(\mathsf{x})\hat\phi_4(\mathsf{x}')\rangle_0
    \right)
    \notag\\
    &+
    \left(
    \langle\normord{\hat\phi_1(\mathsf{x})\hat\phi_4(\mathsf{x}')}\rangle
    +\langle\hat\phi_1(\mathsf{x})\hat\phi_4(\mathsf{x}')\rangle_0
    \right)
    \left(
    \langle\normord{\hat\phi_2(\mathsf{x})\hat\phi_3(\mathsf{x}')}\rangle
    +\langle\hat\phi_2(\mathsf{x})\hat\phi_3(\mathsf{x}')\rangle_0
    \right)
    \notag\\
    &+
    \langle\normord{\hat\phi_1(\mathsf{x})\hat\phi_2(\mathsf{x})}\rangle
    \langle\normord{\hat\phi_3(\mathsf{x}')\hat\phi_4(\mathsf{x}'')}\rangle.
\end{align}
Finally, again using the definition of normal ordering, results in
\begin{align}
    \langle
    \normord{\hat\phi_1(\mathsf{x})\hat\phi_2(\mathsf{x})}
    \normord{\hat\phi_3(\mathsf{x}')\hat\phi_4(\mathsf{x}')}
    \rangle
    =
    \langle\normord{\hat\phi_1(\mathsf{x})\hat\phi_2(\mathsf{x})}\rangle
    \langle\normord{\hat\phi_3(\mathsf{x}')\hat\phi_4(\mathsf{x}')}\rangle
    &+
    \langle\hat\phi_1(\mathsf{x})\hat\phi_3(\mathsf{x}')\rangle
    \langle\hat\phi_2(\mathsf{x})\hat\phi_4(\mathsf{x}')\rangle
    \notag\\
    &+
    \langle\hat\phi_1(\mathsf{x})\hat\phi_4(\mathsf{x}')\rangle
    \langle\hat\phi_2(\mathsf{x})\hat\phi_3(\mathsf{x}')\rangle,
\end{align}
which proves the lemma.\hfill$\square$

We will now show how to obtain Eq.~\eqref{eq:dE/dt_PQG_2}. Inserting $\hat T_{\mu\nu}=\partial_\mu\hat\phi\partial_\nu\hat\phi-\frac{1}{2}\eta_{\mu\nu}(\partial_\alpha\hat\phi)^2$ into $\langle\normord{\dot{\hat{T}}_{kl}(\mathsf{x}')}\normord{\dot{\hat{T}}_{mn}(\mathsf{x}'')}\rangle$ gives
\begin{align}
    \langle\normord{\dot{\hat{T}}_{kl}(\mathsf{x}')}\normord{\dot{\hat{T}}_{mn}(\mathsf{x}'')}\rangle
    =
    \langle
    \normord{
    \left(
    \dot{\hat\phi}'_k\hat\phi'_l+
    \hat\phi'_k\dot{\hat\phi}'_l-
    \eta_{kl}\dot{\hat\phi}'_\alpha\hat\phi'^\alpha
    \right)
    }
    \normord{
    \left(
    \dot{\hat\phi}''_m\hat\phi''_n+
    \hat\phi''_m\dot{\hat\phi}''_n-
    \eta_{mn}\dot{\hat\phi}''_\beta\hat\phi''^\beta
    \right)
    }
    \rangle,
\end{align}
where $\phi'_\alpha=\frac{\partial \hat\phi(\mathsf{x}')}{\partial x^\alpha}$. Expanding out the product and using the lemma gives
\begin{align}
    \langle\normord{\dot{\hat{T}}_{kl}(\mathsf{x}')}\normord{\dot{\hat{T}}_{mn}(\mathsf{x}'')}\rangle
    =
    &
    \langle\normord{\dot\phi'_k\phi'_l}\rangle
    \langle\normord{\dot\phi''_m\phi''_n}\rangle
    +
    \langle\dot\phi'_k\dot\phi''_m\rangle
    \langle\phi'_l\phi''_n\rangle
    +
    \langle\dot\phi'_k\phi''_n\rangle
    \langle\phi'_l\dot\phi''_m\rangle
    \notag\\
    &+
    \langle\normord{\dot\phi'_k\phi'_l}\rangle
    \langle\normord{\phi''_m\dot\phi''_n}\rangle
    +
    \langle\dot\phi'_k\phi''_m\rangle
    \langle\phi'_l\dot\phi''_n\rangle
    +
    \langle\dot\phi'_k\dot\phi''_n\rangle
    \langle\phi'_l\phi''_m\rangle
    \notag\\
    &-\eta_{mn}\left(
    \langle\normord{\dot\phi'_k\phi'_l}\rangle
    \langle\normord{\dot\phi''_\beta\phi''^\beta}\rangle
    +
    \langle\dot\phi'_k\phi''_\beta\rangle
    \langle\phi'_l\dot\phi''^\beta\rangle
    +
    \langle\dot\phi'_k\dot\phi''^\beta\rangle
    \langle\phi'_l\phi''_\beta\rangle
    \right)
    \notag\\
    &+
    \langle\normord{\phi'_k\dot\phi'_l}\rangle
    \langle\normord{\dot\phi''_m\phi''_n}\rangle
    +
    \langle\phi'_k\dot\phi''_m\rangle
    \langle\dot\phi'_l\phi''_n\rangle
    +
    \langle\phi'_k\phi''_n\rangle
    \langle\dot\phi'_l\dot\phi''_m\rangle
    \notag\\
    &+
    \langle\normord{\phi'_k\dot\phi'_l}\rangle
    \langle\normord{\phi''_m\dot\phi''_n}\rangle
    +
    \langle\phi'_k\phi''_m\rangle
    \langle\dot\phi'_l\dot\phi''_n\rangle
    +
    \langle\phi'_k\dot\phi''_n\rangle
    \langle\dot\phi'_l\phi''_m\rangle
    \notag\\
    &-\eta_{mn}\left(
    \langle\normord{\phi'_k\dot\phi'_l}\rangle
    \langle\normord{\dot\phi''_\beta\phi''^\beta}\rangle
    +
    \langle\phi'_k\phi''_\beta\rangle
    \langle\dot\phi'_l\dot\phi''^\beta\rangle
    +
    \langle\phi'_k\dot\phi''^\beta\rangle
    \langle\dot\phi'_l\phi''_\beta\rangle
    \right)
    \notag\\
    &-\eta_{kl}\left(
    \langle\normord{\dot\phi'_\alpha\phi'^\alpha}\rangle
    \langle\normord{\dot\phi''_m\phi''_n}\rangle
    +
    \langle\dot\phi'_\alpha\dot\phi''_m\rangle
    \langle\phi'^\alpha\phi''_n\rangle
    +
    \langle\dot\phi'_\alpha\phi''_n\rangle
    \langle\phi'^\alpha\dot\phi''_m\rangle
    \right)
    \notag\\
    &-\eta_{kl}\left(
    \langle\normord{\dot\phi'_\alpha\phi'^\alpha}\rangle
    \langle\normord{\phi''_m\dot\phi''_n}\rangle
    +
    \langle\dot\phi'_\alpha\phi''_m\rangle
    \langle\phi'^\alpha\dot\phi''_n\rangle
    +
    \langle\dot\phi'_\alpha\dot\phi''_n\rangle
    \langle\phi'^\alpha\phi''_m\rangle
    \right)
    \notag\\
    &+\eta_{kl}\eta_{mn}\left(
    \langle\normord{\dot\phi'_\alpha\phi'^\alpha}\rangle
    \langle\normord{\dot\phi'_\beta\phi'^\beta}\rangle
    +
    \langle\dot\phi'_\alpha\dot\phi''_\beta\rangle
    \langle\phi'^\alpha\phi''^\beta\rangle
    +
    \langle\dot\phi'_\alpha\phi''^\beta\rangle
    \langle\phi'^\alpha\dot\phi''_\beta\rangle
    \right).
\end{align}
where for notational simplicity we omitted the hats on the field operators. Notice that the first terms in each line can be factored together to give $\langle\normord{\dot{\hat{T}}_{kl}(\mathsf{x}')}\rangle\langle\normord{\dot{\hat{T}}_{mn}(\mathsf{x}'')}\rangle$. However, since the expectation value of the stress tensor is time independent in a thermal state, this term vanishes identically. (Note that this is just the SCG term, as seen in Eq.~\eqref{eq:dE/dt_SCG}.) Hence we finally obtain
\begin{align}
    \langle\normord{\dot{\hat{T}}_{kl}(\mathsf{x}')}\normord{\dot{\hat{T}}_{mn}(\mathsf{x}'')}\rangle
    =
    &
    \langle\dot\phi'_k\dot\phi''_m\rangle
    \langle\phi'_l\phi''_n\rangle
    +
    \langle\dot\phi'_k\phi''_n\rangle
    \langle\phi'_l\dot\phi''_m\rangle
    \notag
    +
    \langle\dot\phi'_k\phi''_m\rangle
    \langle\phi'_l\dot\phi''_n\rangle
    +
    \langle\dot\phi'_k\dot\phi''_n\rangle
    \langle\phi'_l\phi''_m\rangle
    \notag\\
    &
    -\eta_{mn}\left(
    \langle\dot\phi'_k\phi''_\beta\rangle
    \langle\phi'_l\dot\phi''^\beta\rangle
    +
    \langle\dot\phi'_k\dot\phi''^\beta\rangle
    \langle\phi'_l\phi''_\beta\rangle
    \right)
    +
    \langle\phi'_k\dot\phi''_m\rangle
    \langle\dot\phi'_l\phi''_n\rangle
    +
    \langle\phi'_k\phi''_n\rangle
    \langle\dot\phi'_l\dot\phi''_m\rangle
    \notag\\
    &+
    \langle\phi'_k\phi''_m\rangle
    \langle\dot\phi'_l\dot\phi''_n\rangle
    +
    \langle\phi'_k\dot\phi''_n\rangle
    \langle\dot\phi'_l\phi''_m\rangle
    -\eta_{mn}\left(
    \langle\phi'_k\phi''_\beta\rangle
    \langle\dot\phi'_l\dot\phi''^\beta\rangle
    +
    \langle\phi'_k\dot\phi''^\beta\rangle
    \langle\dot\phi'_l\phi''_\beta\rangle
    \right)
    \notag\\
    &-\eta_{kl}\left(
    \langle\dot\phi'_\alpha\dot\phi''_m\rangle
    \langle\phi'^\alpha\phi''_n\rangle
    +
    \langle\dot\phi'_\alpha\phi''_n\rangle
    \langle\phi'^\alpha\dot\phi''_m\rangle
    \right)
    -\eta_{kl}\left(
    \langle\dot\phi'_\alpha\phi''_m\rangle
    \langle\phi'^\alpha\dot\phi''_n\rangle
    +
    \langle\dot\phi'_\alpha\dot\phi''_n\rangle
    \langle\phi'^\alpha\phi''_m\rangle
    \right)
    \notag\\
    &+\eta_{kl}\eta_{mn}\left(
    \langle\dot\phi'_\alpha\dot\phi''_\beta\rangle
    \langle\phi'^\alpha\phi''^\beta\rangle
    +
    \langle\dot\phi'_\alpha\phi''^\beta\rangle
    \langle\phi'^\alpha\dot\phi''_\beta\rangle
    \right),
\end{align}
which, when inserted into Eq.~\eqref{eq:dE/dt_PQG}, is of the schematic form given in Eq.~\eqref{eq:dE/dt_PQG_2}. Importantly, notice that all of the two-point correlators are of the form $\langle\hat\phi_1(\mathsf{x}')\hat\phi_2(\mathsf{x}'')\rangle$ and \textit{not} $\langle\hat\phi_1(\mathsf{x}')\hat\phi_2(\mathsf{x}')\rangle$ or $\langle\hat\phi_1(\mathsf{x}'')\hat\phi_2(\mathsf{x}'')\rangle$. This is important because while the latter two expressions are always divergent, $\langle\hat\phi_1(\mathsf{x}')\hat\phi_2(\mathsf{x}'')\rangle$ is only divergent when $\mathsf{x}'$ and $\mathsf{x}''$ are in null separation. As we see in the next section, these light cone divergences can be consistently regularized.

\section{Regularization of passive quantum gravity integrals}
\label{app:regularization}

Eq.~\eqref{eq:dE/dt_PQG_2} gives the power radiated into GWs by a spherical volume $V$ of radius $L$. By spherical symmetry, we can compute this quantity by evaluating Eq.~\eqref{eq:dE/dt_PQG_2} at any point $\bm x$ such that $|\bm x|\gg L$. Therefore let us set $\bm x = (0,0,|\bm x|)=|\bm x|\hat z$.

We will now explicitly show how to compute the first term in Eq.~\eqref{eq:dE/dt_PQG_2}, with $k=l=x$ and $m=n=y$. In the full expression we have to sum over all values of $k,l,m,n$ --- see Eq.~\eqref{eq:dE/dt_function_of_T}. The other terms in the sum can be computed analogously. Thus we are interested in computing
\begin{align}
\label{eq:I1(L)_def}
    I_1(L):=
    \frac{G}{V}\!\!
    \int_\mathbb{R}\!\d[3]{\bm x'}\!\!\!
    \int_\mathbb{R}\!\d[3]{\bm x''}\!
    e^{\frac{-2|\bm x'|^2}{L^2}}
    e^{\frac{-2|\bm x''|^2}{L^2}}
    \langle
    \dot{\hat\phi}_x(\mathsf{x}')
    \dot{\hat\phi}_y(\mathsf{x}'')
    \rangle
    \langle
    {\hat\phi}_x(\mathsf{x}')
    {\hat\phi}_y(\mathsf{x}'')
    \rangle.
\end{align}
Note that, as discussed in the text, we are approximating the integrals over the sphere $V$ of radius $L$ by taking the integral over all $\mathbb{R}$ while inserting the Gaussian ``smearing" function of width $L$ to pick out the appropriate region of space. As an order of magnitude estimate, which is all we are after, this is an appropriate approximation. 

Also recall from the text that in Eq.~\eqref{eq:I1(L)_def} $\mathsf{x}'=(t'_\text{ret},\bm x')$ and $\mathsf{x}''=(t''_\text{ret},\bm x'')$, where $t'_\text{ret}=t-|\bm x-\bm x'|$ and $t''_\text{ret}=t-|\bm x-\bm x''|$ are the retarded times of the source points $x'$ and $x''$ with respect to the field point $x=(t,\bm x)$. Since we are assuming that $|\bm x|\gg L$ we find that $t'_\text{ret}\approx t-(\bm x-\bm x')\cdot \hat z$ and $t''_\text{ret}\approx t-(\bm x-\bm x'')\cdot \hat z$. In particular this results in
\begin{equation}
\label{eq:tau}
    \Delta t:=t'_\text{ret}-t''_\text{ret}=(\bm x'-\bm x'')\cdot \hat z,
\end{equation}
which we will soon make use of.

Differentiating the thermal two-point correlator $\langle{\hat\phi}(\mathsf{x}'){\hat\phi}(\mathsf{x}'')\rangle$ given in Eq.~\eqref{eq:correlator}, and evaluating at the retarded times, we obtain
\begin{align}
    \langle
    {\hat\phi}_x(\mathsf{x}')
    {\hat\phi}_y(\mathsf{x}'')
    \rangle
    &=
    -\frac{2}{\pi^2}(x'_1-x''_1)(x'_2-x''_2)
    \sum_{n=-\infty}^\infty
    \frac{1}{
    \left[
    \Delta\bm x^2
    -
    (\Delta t+\ii n \beta)^2
    \right]^3},
    \\
    \langle
    \dot{\hat\phi}_x(\mathsf{x}')
    \dot{\hat\phi}_y(\mathsf{x}'')
    \rangle
    &=
    \frac{12}{\pi^2}(x'_1-x''_1)(x'_2-x''_2)
    \sum_{n=-\infty}^\infty
    \frac{\Delta\bm x^2
    +
    7(\Delta t+\ii n \beta)^2}{
    \left[
    \Delta\bm x^2
    -
    (\Delta t^2+\ii n \beta)^2
    \right]^5},
\end{align}
where $\Delta\bm x = |\bm x'-\bm x''|$ and $\Delta t$ is given by Eq.~\eqref{eq:tau}. Substituting this into Eq.~\eqref{eq:I1(L)_def} and making the change of variables $\bm u = \bm x'-\bm x''$, $\bm v = \bm x' + \bm x''$ we obtain
\begin{align}
    \label{eq:I1(L)_2}
    I_1(L)
    &=
    -\frac{9G}{4 \pi^3 L^3}
    \sum_{n'=-\infty}^\infty
    \sum_{n''=-\infty}^\infty
    \int_\mathbb{R}\d[3]{\bm v}e^\frac{-|\bm v|^2}{L^2}
    \int_\mathbb{R}\d[3]{\bm u}e^\frac{-|\bm u|^2}{L^2}
    u_x^2 u_y^2
    \frac{|\bm u|^2
    +
    7(\bm u\cdot\hat z+\ii n' \beta)^2}{
    \left[
    |\bm u|^2
    -
    (\bm u\cdot\hat z+\ii n' \beta)^2
    \right]^5}
    \frac{1}{
    \left[
    |\bm u|^2
    -
    (\bm u\cdot\hat z+\ii n'' \beta)^2
    \right]^3}
    \notag\\
    &=
    -\frac{9G}{16\sqrt{\pi^5}}
    \sum_{n'=-\infty}^\infty
    \sum_{n''=-\infty}^\infty
    \int_\mathbb{R}\d[3]{\bm u}e^\frac{-|\bm u|^2}{L^2}
    u_x^2 u_y^2
    \frac{|\bm u|^2
    +
    7(\bm u\cdot\hat z+\ii n' \beta)^2}{
    \left[
    |\bm u|^2
    -
    (\bm u\cdot\hat z+\ii n' \beta)^2
    \right]^5}
    \frac{1}{
    \left[
    |\bm u|^2
    -
    (\bm u\cdot\hat z+\ii n'' \beta)^2
    \right]^3},
\end{align}
where in the second line we performed the $\bm v$ integration. Notice that the $L^3$ from this integral cancelled the $L^3$ in the prefactor. 

To proceed with the calculation, let us write $I_1(L)=I_1^{\textsc{vac}\times\textsc{vac}}(L)+I_1^{\textsc{vac}\times\textsc{no}}(L)+I_1^{\textsc{no}\times\textsc{vac}}(L)+I_1^{\textsc{no}\times\textsc{no}}(L)$, where
\begin{align}
    I_1^{\textsc{vac}\times\textsc{vac}}(L)
    &=
    -\frac{9G}{16\sqrt{\pi^5}}
    \int_\mathbb{R}\d[3]{\bm u}e^\frac{-|\bm u|^2}{L^2}
    u_x^2 u_y^2
    \frac{|\bm u|^2
    +
    7(\bm u\cdot\hat z)^2}{
    \left[
    |\bm u|^2
    -
    (\bm u\cdot\hat z)^2
    \right]^5}
    \frac{1}{
    \left[
    |\bm u|^2
    -
    (\bm u\cdot\hat z)^2
    \right]^3},
    \\
    I_1^{\textsc{vac}\times\textsc{no}}(L)
    &=
    -\frac{9G}{16\sqrt{\pi^5}}
    \sum_{n''\neq 0}
    \int_\mathbb{R}\d[3]{\bm u}e^\frac{-|\bm u|^2}{L^2}
    u_x^2 u_y^2
    \frac{|\bm u|^2
    +
    7(\bm u\cdot\hat z)^2}{
    \left[
    |\bm u|^2
    -
    (\bm u\cdot\hat z)^2
    \right]^5}
    \frac{1}{
    \left[
    |\bm u|^2
    -
    (\bm u\cdot\hat z+\ii n'' \beta)^2
    \right]^3},
    \\
    I_1^{\textsc{no}\times\textsc{vac}}(L)
    &=
    -\frac{9G}{16\sqrt{\pi^5}}
    \sum_{n'\neq 0}
    \int_\mathbb{R}\d[3]{\bm u}e^\frac{-|\bm u|^2}{L^2}
    u_x^2 u_y^2
    \frac{|\bm u|^2
    +
    7(\bm u\cdot\hat z+\ii n' \beta)^2}{
    \left[
    |\bm u|^2
    -
    (\bm u\cdot\hat z+\ii n' \beta)^2
    \right]^5}
    \frac{1}{
    \left[
    |\bm u|^2
    -
    (\bm u\cdot\hat z)^2
    \right]^3},
    \\
    I_1^{\textsc{no}\times\textsc{no}}(L)
    &=
    -\frac{9G}{16\sqrt{\pi^5}}
    \sum_{n'\neq 0}
    \sum_{n''\neq 0}
    \int_\mathbb{R}\d[3]{\bm u}e^\frac{-|\bm u|^2}{L^2}
    u_x^2 u_y^2
    \frac{|\bm u|^2
    +
    7(\bm u\cdot\hat z+\ii n' \beta)^2}{
    \left[
    |\bm u|^2
    -
    (\bm u\cdot\hat z+\ii n' \beta)^2
    \right]^5}
    \frac{1}{
    \left[
    |\bm u|^2
    -
    (\bm u\cdot\hat z+\ii n'' \beta)^2
    \right]^3}.
\end{align}
Here, as discussed in the text, we split up $I_1(L)$ based on the singularities of the integrand. $I_1^{\textsc{vac}\times\textsc{vac}}(L)$ is the most singular since it contains two vacuum correlators. $I_1^{\textsc{vac}\times\textsc{no}}(L)$ and $I_1^{\textsc{no}\times\textsc{vac}}(L)$ are less singular since they contain one vacuum correlator and a non-singular normal ordered correlator in their integrands. Lastly $I_1^{\textsc{no}\times\textsc{no}}(L)$ is non-singular. In order to compute $I_1(L)$, we now have to deal with the divergences present in the first three integrals. 

As discussed in the text, the first step is to follow the usual practice (see Refs.~\cite{Ford2001,Ford2002,Ford2003}) of completely discarding the purely vacuum term. In other words we are only interested in the relative difference between how much a thermal state radiates and how much the vacuum radiates. To further simplify the notation, we rewrite the remaining three terms with units fixed so that $\beta=1$:
\begin{align}
    \label{eq:VacNo}
    I_1^{\textsc{vac}\times\textsc{no}}(L)
    &=
    -\frac{9G}{16\sqrt{\pi^5}}
    \sum_{n''\neq 0}
    \int_\mathbb{R}\d[3]{\bm u}e^\frac{-|\bm u|^2}{L^2}
    u_x^2 u_y^2
    \frac{|\bm u|^2
    +
    7(\bm u\cdot\hat z)^2}{
    \left[
    |\bm u|^2
    -
    (\bm u\cdot\hat z)^2
    \right]^5}
    \frac{1}{
    \left[
    |\bm u|^2
    -
    (\bm u\cdot\hat z+\ii n'')^2
    \right]^3},
    \\
    \label{eq:NoVac}
    I_1^{\textsc{no}\times\textsc{vac}}(L)
    &=
    -\frac{9G}{16\sqrt{\pi^5}}
    \sum_{n'\neq 0}
    \int_\mathbb{R}\d[3]{\bm u}e^\frac{-|\bm u|^2}{L^2}
    u_x^2 u_y^2
    \frac{|\bm u|^2
    +
    7(\bm u\cdot\hat z+\ii n')^2}{
    \left[
    |\bm u|^2
    -
    (\bm u\cdot\hat z+\ii n')^2
    \right]^5}
    \frac{1}{
    \left[
    |\bm u|^2
    -
    (\bm u\cdot\hat z)^2
    \right]^3},
    \\
    \label{eq:NoNo}
    I_1^{\textsc{no}\times\textsc{no}}(L)
    &=
    -\frac{9G}{16\sqrt{\pi^5}}
    \sum_{n'\neq 0}
    \sum_{n''\neq 0}
    \int_\mathbb{R}\d[3]{\bm u}e^\frac{-|\bm u|^2}{L^2}
    u_x^2 u_y^2
    \frac{|\bm u|^2
    +
    7(\bm u\cdot\hat z+\ii n')^2}{
    \left[
    |\bm u|^2
    -
    (\bm u\cdot\hat z+\ii n')^2
    \right]^5}
    \frac{1}{
    \left[
    |\bm u|^2
    -
    (\bm u\cdot\hat z+\ii n'')^2
    \right]^3}.
\end{align}
$L$ is now measured in units of $\beta$. Note that since we already set $k_\textsc{b}=\hbar=c=1$, this completely fixes the units. At the end of our calculation we will restore the quantities $\beta$, $k_\textsc{b}$, $\hbar$ and $c$ so that $I_1(L)$ has the correct dimensions of a power density. Note that since there is no way to combine $\beta$, $k_\textsc{b}$, $\hbar$ and $c$ into a dimensionless quantity, this final restoration of units will be unique.

The next step in evaluating the the integrals \eqref{eq:VacNo}-\eqref{eq:NoNo} is to change to polar coordinates $u_x=r\cos\theta$ and $u_y=r\sin\theta$. This gives
\begin{align}
    \label{eq:VacNo_2}
    I_1^{\textsc{vac}\times\textsc{no}}(L)
    &=
    -\frac{9G}{64\sqrt{\pi^3}}
    \sum_{n''\neq 0}
    \int_0^\infty\dif r
    \int_{-\infty}^\infty\dif u_z
    e^\frac{-r^2}{L^2}
    e^\frac{-u_z^2}{L^2}
    \frac{r^2+8u_z^2}{r^5}
    \frac{1}{
    \left[
    r^2+u_z^2
    -
    (u_z+\ii n'')^2
    \right]^3},
    \\
    \label{eq:NoVac_2}
    I_1^{\textsc{no}\times\textsc{vac}}(L)
    &=
    -\frac{9G}{64\sqrt{\pi^3}}
    \sum_{n'\neq 0}
    \int_0^\infty\dif r
    \int_{-\infty}^\infty\dif u_z
    e^\frac{-r^2}{L^2}
    e^\frac{-u_z^2}{L^2}
    \frac{1}{r}
    \frac{r^2+u_z^2
    +
    7(u_z+\ii n')^2}{
    \left[
    r^2+u_z^2
    -
    (u_z+\ii n')^2
    \right]^5},
    \\
    \label{eq:NoNo_2}
    I_1^{\textsc{no}\times\textsc{no}}(L)
    &=
    -\frac{9G}{64\sqrt{\pi^3}}
    \sum_{n'\neq 0}
    \sum_{n''\neq 0}
    \int_0^\infty\dif r
    \int_{-\infty}^\infty\dif u_z
    e^\frac{-r^2}{L^2}
    e^\frac{-u_z^2}{L^2}
    r^5
    \frac{r^2+u_z^2
    +
    7(u_z+\ii n')^2}{
    \left[
    r^2+u_z^2
    -
    (u_z+\ii n')^2
    \right]^5}
    \frac{1}{
    \left[
    r^2+u_z^2
    -
    (u_z+\ii n'')^2
    \right]^3}.
\end{align}
The integral \eqref{eq:NoNo_2} is convergent and can be computed numerically. On the other hand the the integrals Eq.~\eqref{eq:VacNo_2} and \eqref{eq:NoVac_2} are divergent and need regularizing. We will now regulate these integrals via an integration by an parts procedure~\cite{Davies1989,Davies1990}, as is common practice in PQG calculations~\cite{Ford2001,Ford2002,Ford2003}.

Let us start with $I_1^{\textsc{vac}\times\textsc{no}}(L)$ in Eq.~\eqref{eq:VacNo_2}. Note that the integrand has a higher order singularity at $r=0$. In order to regularize this divergence, we first rewrite the $r$ integral to go from $-\infty$ to $\infty$:
\begin{align}
    \label{eq:VacNo_3}
    I_1^{\textsc{vac}\times\textsc{no}}(L)
    &=
    -\frac{9G}{128\sqrt{\pi^3}}
    \sum_{n''\neq 0}
    \int_{-\infty}^\infty\dif r
    \int_{-\infty}^\infty\dif u_z
    e^\frac{-r^2}{L^2}
    e^\frac{-u_z^2}{L^2}
    \frac{r^2+8u_z^2}{|r|^5}
    \frac{1}{
    \left[
    r^2+u_z^2
    -
    (u_z+\ii n'')^2
    \right]^3}.
\end{align}
Next, we insert the identity
\begin{equation}
    \label{eq:magic_identity}
    \frac{1}{|r|^n}
    =
    \frac{(-1)^{n-1}\sgn(r)}{(n-1)!}\frac{\dif^{\,n}}{\dif r^n}\log|r|,
\end{equation}
with $n=5$ into Eq.~\eqref{eq:VacNo_3} to obtain
\begin{align}
    \label{eq:VacNo_4}
    I_1^{\textsc{vac}\times\textsc{no}}(L)
    &=
    -\frac{9G}{128(4!)\sqrt{\pi^3}}
    \sum_{n''\neq 0}
    \int_{-\infty}^\infty\!\!\!\dif r
    \int_{-\infty}^\infty\!\!\!\dif u_z\,
    e^\frac{-r^2}{L^2}
    e^\frac{-u_z^2}{L^2}
    \sgn(r)\frac{\dif^{\,5}}{\dif r^5}
    \log|r|
    \frac{r^2+8u_z^2}{
    \left[
    r^2+u_z^2
    -
    (u_z+\ii n'')^2
    \right]^3}.
\end{align}
Now for the key step: integrate the $r$ integral five times by parts. This gives
\begin{align}
    \label{eq:VacNo_5}
    I_{1,\text{reg}}^{\textsc{vac}\times\textsc{no}}(L)
    &=
    \frac{9G}{128(4!)\sqrt{\pi^3}}
    \sum_{n''\neq 0}
    \int_{-\infty}^\infty\!\!\!\!\!\dif r
    \int_{-\infty}^\infty\!\!\!\!\!\dif u_z\,
    e^\frac{-u_z^2}{L^2}
    \log|r|
    \frac{\dif^{\,5}}{\dif r^5}
    \left[
    e^\frac{-r^2}{L^2}
    \frac{\sgn(r)(r^2+8u_z^2)}{
    \left[
    r^2+u_z^2
    -
    (u_z+\ii n'')^2
    \right]^3}
    \right],
\end{align}
where the subscript ``reg" indicates that we have regularized the integral. Indeed, this integral contains only a logarithmic singularity, which is integrable. We can further simplify the integral if we assume that $L\gg 1$, i.e. $L\gg \beta$. (Note that we have already made this assumption to ensure that the energy of GWs is well defined). In that case, noting that for any differentiable function $f(r)$
\begin{equation}
    \frac{\dif^{\,n}}{\dif r^n}
    \left[
    e^\frac{-r^2}{L^2} f(r)
    \right]
    =
    e^\frac{-r^2}{L^2}
    \frac{\dif^{\,n}}{\dif r^n}f(r)
    +\mathcal{O}\left(\frac{1}{L^2}\right),
\end{equation}
we can neglect the $\mathcal{O}\left(\frac{1}{L^2}\right)$ terms and write
\begin{align}
    \label{eq:VacNo_reg}
    I_{1,\text{reg}}^{\textsc{vac}\times\textsc{no}}(L)
    &=
    \frac{9G}{64(4!)\sqrt{\pi^3}}
    \sum_{n''\neq 0}
    \int_{0}^\infty\!\!\!\!\!\dif r
    \int_{-\infty}^\infty\!\!\!\!\!\dif u_z\,
    e^\frac{-r^2}{L^2}
    e^\frac{-u_z^2}{L^2}
    \log(r)
    \frac{\dif^{\,5}}{\dif r^5}
    \left[
    \frac{r^2+8u_z^2}{
    \left[
    r^2+u_z^2
    -
    (u_z+\ii n'')^2
    \right]^3}
    \right],
\end{align}
where we have also changed the $r$ integral back to the interval $(0,\infty)$. We can regularize $I_1^{\textsc{no}\times\textsc{vac}}(L)$ in the same manner to obtain
\begin{align}
    \label{eq:NoVac_reg}
    I_{1,\text{reg}}^{\textsc{no}\times\textsc{vac}}(L)
    &=
    \frac{9G}{64\sqrt{\pi^3}}
    \sum_{n'\neq 0}
    \int_{0}^\infty\!\!\!\!\!\dif r
    \int_{-\infty}^\infty\!\!\!\!\!\dif u_z\,
    e^\frac{-r^2}{L^2}
    e^\frac{-u_z^2}{L^2}
    \log(r)
    \frac{\dif}{\dif r}
    \left[
    \frac{r^2+u_z^2
    +
    7(u_z+\ii n')^2}{
    \left[
    r^2+u_z^2
    -
    (u_z+\ii n')^2
    \right]^5}
    \right].
\end{align}
Thus, combining Eqs.~\eqref{eq:NoNo_2},~\eqref{eq:VacNo_reg} and \eqref{eq:NoVac_reg}, the regularized version of the integral $I_1(L)$ is
\begin{align}
    \label{eq:I1(L)_reg}
    I_{1,\text{reg}}(L)
    =
    \frac{9G}{64\sqrt{\pi^3}}
    \int_{0}^\infty\!\!\!\!\!\dif r
    \int_{-\infty}^\infty\!\!\!\!\!\dif u_z\,
    e^\frac{-r^2}{L^2}
    e^\frac{-u_z^2}{L^2}
    \Bigg(&
    \frac{\log(r)}{4!}
    \frac{\dif^{\,5}}{\dif r^5}
    \left[
    \sum_{n''\neq 0}
    \frac{r^2+8u_z^2}{
    \left[
    r^2+u_z^2
    -
    (u_z+\ii n'')^2
    \right]^3}
    \right]
    \notag\\
    &+
    \log(r)
    \frac{\dif}{\dif r}
    \left[
    \sum_{n'\neq 0}
    \frac{r^2+u_z^2
    +
    7(u_z+\ii n')^2}{
    \left[
    r^2+u_z^2
    -
    (u_z+\ii n')^2
    \right]^5}
    \right]
    \notag\\
    &-
    \sum_{n'\neq 0}
    \sum_{n''\neq 0}
    \frac{r^2+u_z^2
    +
    7(u_z+\ii n')^2}{
    \left[
    r^2+u_z^2
    -
    (u_z+\ii n')^2
    \right]^5}
    \frac{r^5}{
    \left[
    r^2+u_z^2
    -
    (u_z+\ii n'')^2
    \right]^3}
    \Bigg).
\end{align}
We evaluate this integral numerically in the next section.

\section{Evaluation of passive quantum gravity integrals}
\label{app:evaluation}

\begin{figure}
    \centering
    \includegraphics[width=0.5\textwidth]{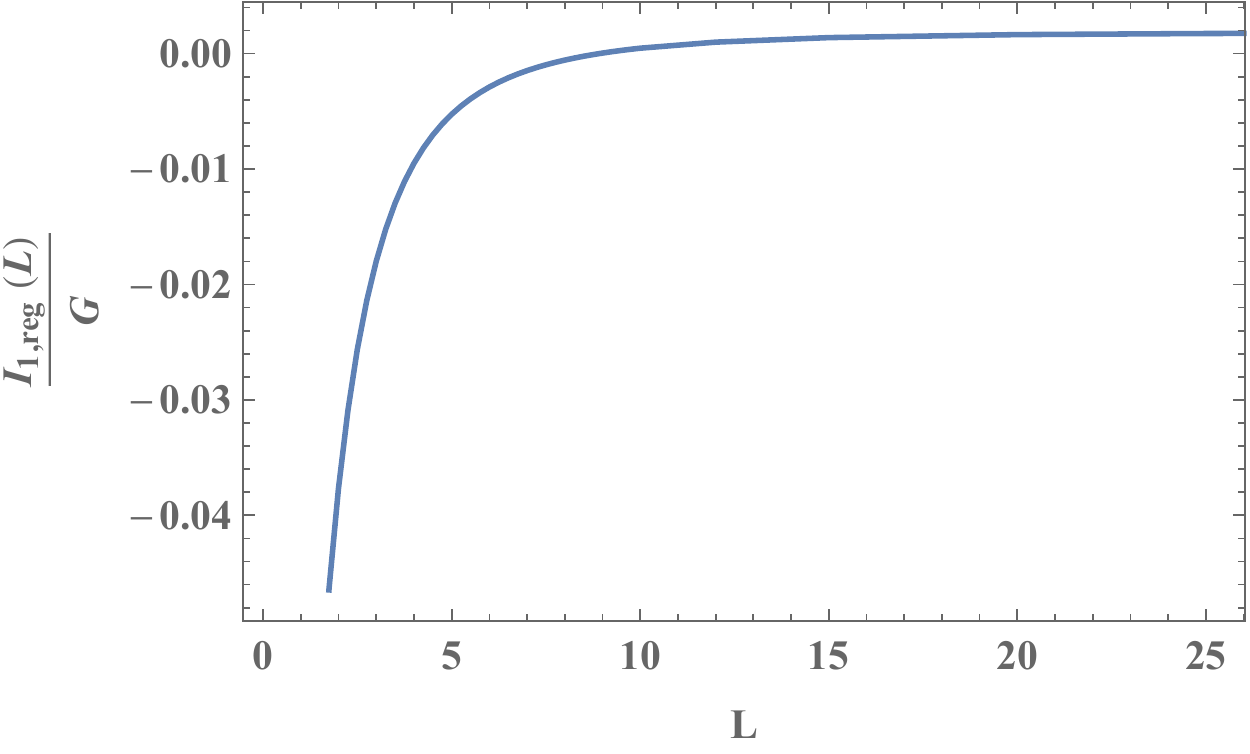}
    \caption{Plot of the regularized contribution $I_{1,\text{reg}}(L)$ to the radiated power density of the CMB, $\frac{\dif \epsilon}{\dif t}$, versus the length scale $L$ of the volume occupied by the CMB. Note that $\displaystyle\lim_{L\rightarrow\infty}I_{1,\text{reg}}(L)\approx 1.81\times10^{-3}$.}
    \label{fig:I1_vs_L}
\end{figure}

Let us now evaluate the regularized integral $I_{1,\text{reg}}(L)$ from Eq.~\eqref{eq:I1(L)_reg}. In particular we are interested in the value of this integral in the $L\gg 1$ (i.e. $L\gg\beta$) limit, since this is the limit that we have made at various points throughout our derivation. Note that, as discussed in the text, we expect the integral to be independent of $L$ in this regime. This is because the function in the round parentheses in Eq.~\eqref{eq:I1(L)_reg} has strong support only if $r\lesssim 1$ and $|u_z|\lesssim 1$, while the Gaussian functions are essentially independent of $L$ in this region as long as $L\gg 1$. In Fig.~\ref{fig:I1_vs_L} we plot $I_{1,\text{reg}}(L)$ versus $L$ and find that this is indeed the case.

In particular we find that $\displaystyle\lim_{L\rightarrow\infty}I_{1,\text{reg}}(L)\approx 1.81\times10^{-3}$, and therefore we can write
\begin{equation}
    I_{1,\text{reg}}(L)\approx
    \left(1.81\times10^{-3}\right)
    \frac{G(k_\textsc{b}T)^7}{\hbar^5 c^8} \text{  for }L\gg\beta,
\end{equation}
where we have restored the scales $\beta=T^{-1}$, $k_\textsc{b}$, $\hbar$ and $c$ in the only possible way that ensures $I_{1,\text{reg}}(L)$ has the units of $\frac{\dif \epsilon}{\dif t}$, i.e. of a power density. Note that, up to a numerical prefactor, this is our expression for $\frac{\dif \epsilon}{\dif t}$ in Eq.~\eqref{eq:dE/dt_PQG_final}. Crucially, the numerical prefactors do not depend on any scales and so they must be close to order unity, and can therefore be ignored in our order of magnitude estimate. (In other words a prefactor of order $10^{-3}$ is completely unimportant when we note that, even if $T=3000$ K, the CMB temperature at the time of matter-radiation decoupling, $\frac{G(k_\textsc{b}T)^7}{\hbar^5 c^8}$ is of order $10^{-42}$). Hence
\begin{equation}
    \frac{\dif \epsilon}{\dif t}
    \Big|_\textsc{pqg}
    \sim
    \frac{G(k_\textsc{b}T)^7}{\hbar^5 c^8},
\end{equation}
is roughly the GW power density radiated by a CMB volume of length scale $L\gg \beta$.

\twocolumngrid

\bibliography{references}
\bibliographystyle{apsrev4-1}

\end{document}